\documentclass[
aps,
pra,
showpacs
amsmath,
amssymb,
onecolumn,
floatfix,
longbibliography,
letterpaper,
lengthcheck,
superscriptaddress,
]{revtex4-2}

\usepackage{graphicx}
\usepackage{booktabs}
\usepackage{floatflt,epsfig} 
\usepackage{mathtools}
\usepackage{color}
\usepackage{braket}
\usepackage[normalem]{ulem}
\usepackage[english]{babel}
\usepackage{blindtext}
\usepackage{lipsum}  
\usepackage{amsthm}
\usepackage{tabularx}
\usepackage{bm}
\usepackage{dsfont}
\usepackage{bbold}
\usepackage{ulem}
\usepackage{mwe,tikz}
\usepackage[percent]{overpic}
\usepackage{todonotes}
\usepackage{algorithm}
\usepackage{multirow}
\usepackage{algpseudocode}
\usepackage{csquotes}
\usepackage{bbold}
\usepackage{soul}
\usepackage{xcolor}
\usepackage{url}

\usepackage[colorlinks]{hyperref}
\hypersetup{%
	plainpages=true,
	breaklinks=true,
	hypertexnames=false,
	pageanchor=true,
	colorlinks=true,
	linkcolor={blue},
	citecolor={blue},
	urlcolor={blue},
	anchorcolor={black}
}

\newcommand{\figref}[1]{\mbox{Fig.~\ref{#1}}}
\newcommand{\tabref}[1]{\mbox{Table~\ref{#1}}}

\renewcommand{\eqref}[1]{\mbox{Eq.~(\ref{#1})}}

\begin{document}

\title{RydbergGPT}

\author{David Fitzek}
\affiliation{Department of Microtechnology and Nanoscience, Chalmers University of Technology, 412 96 Gothenburg, Sweden}
\affiliation{Volvo Group Trucks Technology, 405\,08 Gothenburg, Sweden}
\author{Yi Hong Teoh}
\affiliation{Department of Physics and Astronomy, University of Waterloo, 200 University Ave. West, Waterloo, Ontario N2L 3G1, Canada}
\author{Hin Pok Fung}
\affiliation{Department of Physics and Astronomy, University of Waterloo, 200 University Ave. West, Waterloo, Ontario N2L 3G1, Canada}
\author{Gebremedhin A. Dagnew}
\affiliation{Department of Physics and Astronomy, University of Waterloo, 200 University Ave. West, Waterloo, Ontario N2L 3G1, Canada}
\author{Ejaaz Merali}
\affiliation{Department of Physics and Astronomy, University of Waterloo, 200 University Ave. West, Waterloo, Ontario N2L 3G1, Canada}
\author{M. Schuyler Moss}
\affiliation{Department of Physics and Astronomy, University of Waterloo, 200 University Ave. West, Waterloo, Ontario N2L 3G1, Canada}
\author{Benjamin MacLellan}
\affiliation{Department of Physics and Astronomy, University of Waterloo, 200 University Ave. West, Waterloo, Ontario N2L 3G1, Canada}
\author{Roger G. Melko}
\affiliation{Department of Physics and Astronomy, University of Waterloo, 200 University Ave. West, Waterloo, Ontario N2L 3G1, Canada}
\affiliation{Perimeter Institute for Theoretical Physics, Waterloo, Ontario N2L 2Y5, Canada}

\begin{abstract}
We introduce a generative pretained transformer (GPT) designed to learn the measurement outcomes of a neutral atom array quantum computer.
Based on a vanilla transformer, our encoder-decoder architecture takes as input the interacting Hamiltonian, and outputs an autoregressive sequence of qubit measurement probabilities.
Its performance is studied in the vicinity of a quantum phase transition in Rydberg atoms in a square lattice array.
We explore the ability of the architecture to generalize, by producing groundstate measurements for Hamiltonian parameters not seen in the training set.
We focus on examples of physical observables obtained from inference on three different models, trained in fixed compute time on a single NVIDIA A100 GPU.
These can act as benchmarks for the scaling of larger RydbergGPT models in the future.
Finally, we provide  RydbergGPT open source, to aid in the development of  foundation models based off of a wide variety of quantum computer interactions and data sets in the future.
\end{abstract}

\date{\today}
\maketitle

\section{Introduction}

Generative models have emerged as a key technology for predicting the probabilistic behavior of a quantum system.
Their most basic task is to produce a target sequence representing qubit measurement outcomes; e.g.~projective measurements distributed according to the Born rule of quantum mechanics.
A wide variety of architectures have been explored for this purpose, from restricted Boltzmann machines~\cite{Nomura2017, Gao2017, Carleo2017, Torlai2018, Carrasquilla2021, Viteritti2022, Torlai2019, Carrasquilla2019Mar, Czischek2018}, to recurrent neural networks~\cite{Moss2024, Czischek2022, Carrasquilla2017a, Carrasquilla2021, Morawetz2021, Hibat_Allah2023, Hibat_Allah_2021, hibatallah2024supplementing} and transformers~\cite{Wang2022PredictingPropertiesQuantum, Carrasquilla2021Probabilisticsimulationquantum, Zhang_2023, Sprague2024, Cha2022a, Viteritti2023, viteritti2024transformer,Lange2024}.
While the above models have been trained to output target sequences drawn from a single parameterized distribution, the utility of generative models can be expanded by adopting modern encoder-decoder architectures.
In this case, the encoder takes a source sequence and maps it to a context vector.
This is then passed to the decoder, which combines it with other inputs and maps it to the target sequence.

The most powerful architecture to emerge recently is the attention-based transformer~\cite{Vaswani2017Attentionallyou}, variations of which underlie most of the large language models (LLMs) being developed by academia and industry today~\cite{zhang2023complete, gozalobrizuela2023chatgpt, Wu2023, Brown2020Languagemodelsare, Touvron2023LLaMAOpenEfficient, nakaji2024generative}.
A key challenge in physics that can be tackled by the vanilla encoder-decoder transformer structure is the task of predicting the output of a quantum computer given its experimental settings.
In the simplest example, the source sequence is a description of an interacting Hamiltonian, and the target sequence is the conditional probability of each qubit measurement outcome, selected from a dictionary of size two.
Viewing the task as a language problem, the self-attention mechanism encodes context, or correlations, between qubit measurement outcomes. 
Self-attention has been shown to be a suitable mechanism of encoding the possible long-range correlations -- including entanglement -- found in quantum systems without the need to pass hidden or latent vectors along the sequence \cite{Zhang_2023, Wang2022PredictingPropertiesQuantum, Sprague2024}. 
The impressive demonstrations of scaling in large language models gives hope that the transformer architecture will benefit quantum computers as they continue to grow in size and capability~\cite{kaplan2020scaling}.

In this work, we train an attention-based transformer to learn the distribution of qubit measurement outcomes corresponding to a Hamiltonian describing an array of interacting Rydberg atoms.
Using the encoder-decoder architecture, we use simulated Rydberg occupation data drawn from a two-dimensional qubit array to train three different models.  
Once trained, each model can produce new data via autoregressive sampling of the decoder.
In the case of a Rydberg atom array in its groundstate, the autoregressive output can be interpreted as the complete quantum wavefunction, allowing for the calculation 
of many physical observables of interest.
We report results for estimators of the energy and other physical observables.
By varying the Hamiltonian that is input to the encoder, we find that the transformer is able to accurately predict qubit measurement outcomes in regions outside of the training regime.
% The robustness of groundstate inference to thermal noise in the training data suggest that the architecture is suitable to be trained from real experimental data in the future.
Our results give optimism that attention-based transformer models are well-positioned to provide a powerful approach for predicting the properties of quantum many-body systems.

\section{Rydberg atom array physics}

Rydberg atoms arrays are a powerful platform for quantum information processing where interacting qubits are encoded with the electronic ground state $| g \rangle$ and excited (Rydberg) state $| r \rangle$ for each atom~\cite{Wu2021concisereviewRydberg, Lukin2001DipoleBlockadeQuantum, Browaeys2020Manybodyphysicsindividually, Jaksch2000FastQuantumGates, Endres2016Atombyatomassemblydefectfree}. 
We consider a system of $N=L \times L$ atoms arranged on a square lattice.
The governing Hamiltonian defining the Rydberg atom array interactions has the following form:
\begin{align}
    \hat{H} &= \sum_{i<j} \frac{C_6}{\lVert \mathbf{r}_i - \mathbf{r}_j \rVert^6} \hat{n}_i \hat{n}_j -\delta \sum_{i=1}^N \hat{n}_i - \frac{\Omega}{2} \sum_{i=1}^N \hat{\sigma}^x_i, \label{eq:rydberg_hamiltonian} \\
    C_6 &= \Omega \left( \frac{R_b}{a} \right)^6 \label{eq:rydberg_bloqade}, \quad
    V_{ij} =  \frac{a^6}{\lVert \mathbf{r}_i - \mathbf{r}_j \rVert^6},
\end{align}
where $\hat{\sigma}^{x}_{i} = \vert g \rangle_i \langle r\vert_i + \vert r \rangle_i \langle g\vert_i $, the occupation number operator $\hat{n}_i = \frac{1}{2} \left( \hat{\sigma}_{i} + \mathbb{1} \right) =  \vert r\rangle_i \langle r \vert_i$ and $\hat{\sigma}_{i} = \vert r \rangle_i \langle r \vert_i - \vert g \rangle_i \langle g \vert_i$.
The experimental settings of a Rydberg atom array are controlled by the detuning from resonance $\delta$, Rabi frequency $\Omega$, lattice length scale $a$ and the positions of the atoms $\{\mathbf{r}_i\}_i^N$.
From \eqref{eq:rydberg_bloqade}, we obtain a symmetric matrix $\mathbf{V}$, that encapsulates the relevant information about the lattice geometry, and derive the Rydberg blockade radius $R_b$, within which simultaneous excitations are penalized~\cite{Wu2021concisereviewRydberg, Lukin2001DipoleBlockadeQuantum}.
Finally, for the purposes of our study, the atom array is considered to be affected by thermal noise, in equilibrium at a temperature $T = 1/\beta$. The experimental settings are thus captured by the set of parameters $\mathbf{x} = (\Omega, \delta/\Omega, R_b/a, \mathbf{V}, \beta \Omega)$. 
By adjusting the free parameters in the Rydberg Hamiltonian, the system can be prepared in various phases of matter, separated by distinct phase transitions~\cite{Endres2016Atombyatomassemblydefectfree, Ebadi2021Quantumphasesmatter, Xu2021FastPreparationDetection, Samajdar_2021, Miles2023, Kalinowski2022}.

Rydberg atom array experiments are able to perform projective measurements of the qubit states in the occupation basis. These measurements are informationally complete at $T=0$ due to the positive real-valued ground state wavefunction \cite{bravyi2007complexity, Endres2016Atombyatomassemblydefectfree, Merali2023StochasticSeriesExpansiona, Xu2021}. 
In the next section, we consider these projective measurements as training vectors for an unsupervised learning strategy for our generative model.  The goal of the generative model after training will be to predict, given a Hamiltonian, the distribution of projective measurements in the quantum system.

\section{Transformer architecture} \label{sec:nn_wavefunction}

\begin{figure}[ht]
  \centering
  \includegraphics[width=0.48\textwidth]{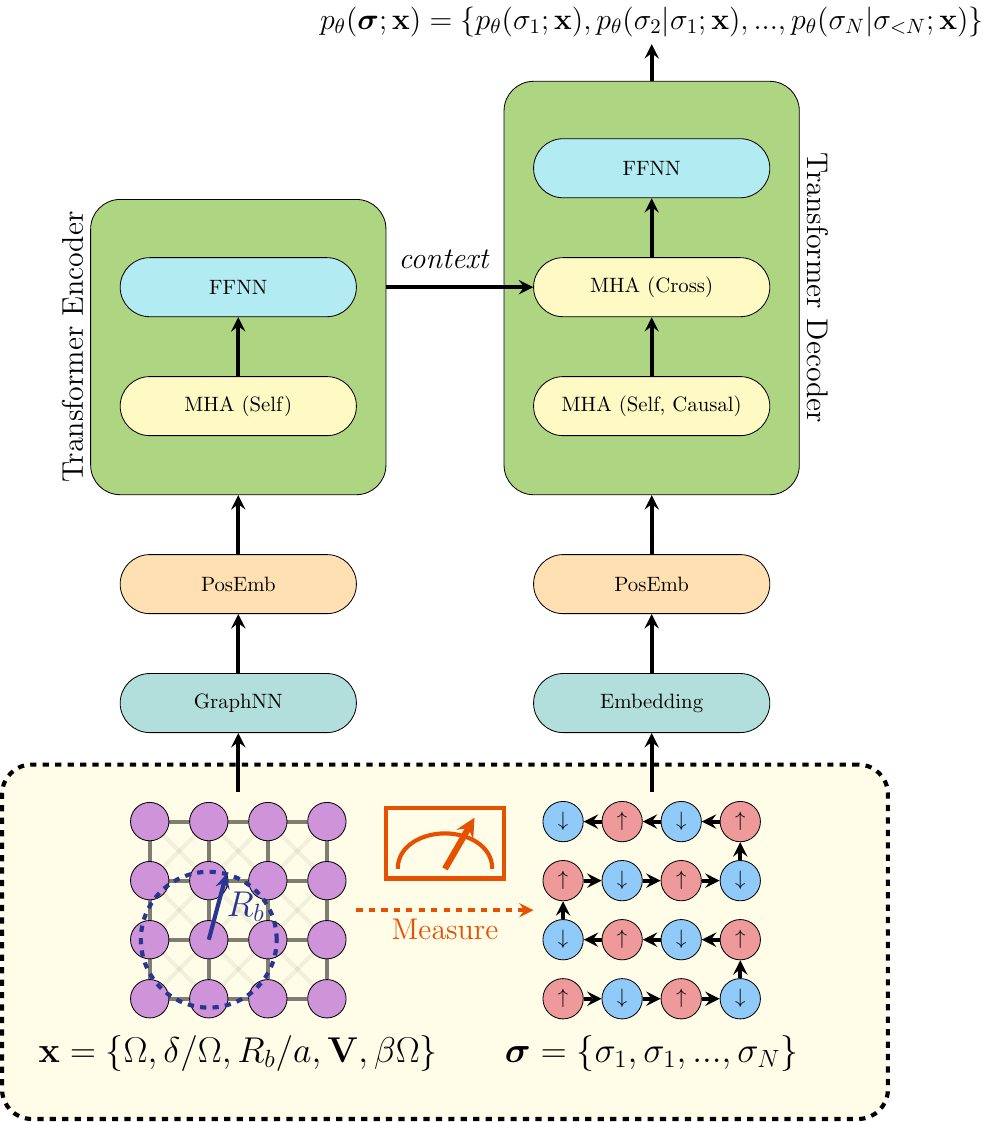}
  \caption{Overview of RydbergGPT.
  The bottom section represents the data from the physical system used in the architecture.
  On the left is a diagram representing the experimental settings $\mathbf{x}$ of the system and on the right is a corresponding projective measurement $\boldsymbol{\sigma}$.
  The experimental settings are inputs to the encoder portion of the model, composed of a graph neural network and the transformer encoder, and is processed to form a context vector.
  This context vector is combined with the sample to obtain a chain of conditional probabilities from the transformer decoder, the product of which is the probability of obtaining the sample from the system.
  The sequence of conditional probabilities is mapped onto the 2D lattice with a ``snake'' path as illustrated.
  }
  \label{fig:model_architecture}
\end{figure}

In order to learn the qubit measurement distribution of the Rydberg atom array, we employ the transformer encoder-decoder architecture \cite{Vaswani2017Attentionallyou}.
These generative models are universal sequence-to-sequence function approximators and, granted sufficient tunable parameters $\theta$, can encode arbitrarily complex functions~\cite{Yun2020}.
Transformers belong to the class of generative models known as autoregressive models.
Autoregressive models learn conditional probability distributions over the $N$ individual variables, such as qubit states $\boldsymbol{\sigma} = \{\sigma_i\}_i^N$, $p_{\theta}(\sigma_i | \sigma_{i-1}, \ldots, \sigma_1)$. 
The joint distribution that arises from these conditionals can be used to represent the likelihood of measurement outcomes,
\begin{align}
 p_{\theta}(\boldsymbol{\sigma}) = \prod_{i=1}^n p_{\theta}\left(\sigma_i \mid \sigma_{i-1}, \ldots, \sigma_1\right).\label{eq:cond_prob_distr}
\end{align}
This is the decoder output, which can be sampled to infer any number of qubit measurements.
This output can be related to a pure quantum wavefunction for example through the Born rule $p_{\theta}(\boldsymbol{\sigma}) = |\Psi_{\theta}(\boldsymbol{\sigma})|^2$.
Additionally, under the assumption of a positive real-valued wavefunction, the transformer's output captures the full quantum state, i.e. $\Psi_{\theta}(\boldsymbol{\sigma}) = \sqrt{p_{\theta}(\boldsymbol{\sigma})}$~\cite{Moss2024, Wu2021concisereviewRydberg}, where $\theta$ denotes the model parameters.

An encoder that maps the experimental Hamiltonian settings to a context vector can be used to condition this decoder.
The encoder-decoder architecture that we use is illustrated in~\figref{fig:model_architecture} and follows the overall design introduced by Vaswani et al.~\cite{Vaswani2017Attentionallyou}.

Note that the source sequence for our encoder, the experimental Hamiltonian settings $\mathbf{x}$, have a natural graph representation.
The interactions $\mathbf{V}$ can be considered as the edge attributes and the remaining variables node attributes.
The field of machine learning has proposed a multitude of models for processing such graphs~\cite{Kipf2017Semisupervisedclassificationgraph}.
In our architecture, we choose to process the graph with a convolutional graph neural network~\cite{Kipf2017Semisupervisedclassificationgraph, Wang2019gcn}.
This allow us to map the graph, which scales quadratically  in the number of atoms $N$, to a sequence representation that scales linearly.
To further capture correlations, we feed the sequence to the vanilla transformer encoder, consisting of multiple self multi-head attention and feedforward blocks, and obtain the context vector.
The decoder autoregressively parameterizes the equilibrium state of the system with respect to the context via an attention mechanism between the context and the intermediate state within the decoder.
The choice of hyperparameters for the architecture is outlined in \tabref{tab:hyperparams}.

\begin{table}[ht]
    \caption{The transformer encoder-decoder architecture and training parameters. Including the dimension of the feedforward network layer ($d_{\mathrm{ff}}$), the model dimension ($d_{\mathrm{model}}$), the graph dimension ($d_{\mathrm{graph}}$) as well as the total number of trainable parameters.}
    \begin{tabular}{l|l}
        \toprule
        \textbf{Parameter} & \textbf{Value} \\ 
        \midrule
        \multicolumn{2}{c}{Neural network architecture} \\        
        \midrule
        $d_{\mathrm{ff}}$ & 128 \\
        $d_{\mathrm{model}}$ & 32 \\
        $d_{\mathrm{graph}}$ & 64 \\
        num heads & 8 \\
        num blocks encoder & 1 \\
        num blocks decoder & 3 \\
        num graph layers & 2 \\
        trainable params & 66562 \\
        \midrule
        \multicolumn{2}{c}{Training hyperparameters} \\
        \midrule
        batch size & 1024\\
        optimizer & AdamW \\
        dropout & 0.1 \\
        learning rate & 0.001 \\
        learning rate schedule & Cosine annealing warm start \\
        $T_0$ & 1 \\
        $T_{\mathrm{mult}}$ & 2 \\
        $\eta_{\mathrm{min}}$ & 0.00001 \\
        dataset buffer & 50 \\ 
        \bottomrule
    \end{tabular}
    \label{tab:hyperparams}
\end{table}

A powerful property of autoregressive models is the ability to perform independent and identically distributed sampling without the use of Markov chain Monte Carlo (MCMC).
To perform sampling, we input the experimental settings of interest to our encoder and obtain the context vector as output.
Following that, the decoder, conditioned on this context, takes as input a starting token and samples the first element of the sequence.
Then, we iteratively sample the $i^{\mathrm{th}}$ token conditioned on all tokens preceding it.
We traverse the lattice of atoms systematically with the snake path convention, shown in the bottom right of Fig.~\ref{fig:model_architecture}.

\section{Training transformer models} \label{sec:training}

We follow a data-driven training procedure, where we train transformers on a synthetic dataset 
$\mathcal{D}$ composed of pairs of source and target sequences.
The former corresponds to experimental settings $\mathbf{x}$, input to the encoder, and the latter corresponds to binary qubit measurement data, input to the decoder.
Our target sequences are simulated Rydberg occupation measurements, produced via 
Stochastic Series Expansion quantum Monte Carlo (QMC) method described in Ref.~\cite{Merali2023StochasticSeriesExpansiona}. 

We employ our QMC to produce uncorrelated samples, which we collect across a wide range of settings.
Specifically, we choose values of the inverse temperature $\beta \Omega$ within the set $\{ 0.5,1,2,4,8,16 \}$, the detuning parameter $\delta / \Omega $ over the range $[-0.364, 3.173]$, the Rydberg blockade $R_b / a$ within $\{ 1.05,1.15,1.3 \}$, and the linear system size $L$ for the values $\{ 5,6,11,12,15,16 \}$.
Without loss of generality, we take $\Omega = 1$ henceforth.
For each configuration, we generate a dataset containing $\lvert \mathcal{D} \rvert = 10^5$ entries, each corresponding to a binary measurement sampled from the QMC simulation.
The datasets are selected to explore $\delta/\Omega$ 
values near a quantum phase transition in the square-lattice Rydberg atom arrays, which occurs at approximately $\delta/\Omega = 1.1$~\cite{Merali2023StochasticSeriesExpansiona}.

A transformer architecture with parameters $\theta$ is trained in an unsupervised manner with a dataset $\mathcal{D}$ by minimizing the Kullback-Leibler (KL) divergence, resulting in the loss function:
\begin{equation}
\mathcal{L}(\theta) \approx -\frac{1}{|\mathcal{D}|} \sum_{\boldsymbol{\sigma} \in \mathcal{D}} \ln p_{\theta}(\boldsymbol{\sigma}). \label{eq:nll}
\end{equation}
This loss is minimized when the decoder output and the target distributions are equal. 
We initialize each model using the method introduced by Glorot \textit{et al.}~\cite{Glorot2010}
We use the AdamW optimizer~\cite{Loshchilov2019decoupled}, which is an extension to stochastic gradient descent with an adaptive learning rate and weight regularization, in combination with a cosine annealing warm restart learning rate scheduler~\cite{loshchilov2017sgdr}.
Details for the training hyperparameters can be found in \tabref{tab:hyperparams}. 

In this manuscript, we train three models with different subsets of the simulated data.
Specifically, the models are all trained on data at $R_b / a = 1.15$ and across the values for $\delta/\Omega$ and $\beta \Omega$.
However, each model is trained with different sets of linear size $L$.
Model $M_1$ is trained with data for systems of size $L=5,6$, model $M_2$ with $L=5,6,11,12$ and model $M_3$ with $L=5,6,11,12,15,16$, refer to \tabref{tab:measurements_lattice_sizes} for the total number of tokens, i.e. measurement outcome of a single qubit, for each lattice size.
As we generate $10^5$ samples for each lattice size, the larger lattice sizes have more tokens compared to the smaller ones.
\begin{table}
    \caption{
    Number of tokens, in  units of \(10^8\), contained in our dataset for various sizes $L$.
    Each token corresponds to the measurement outcome of a single Rydberg qubit.
    For each configuration, i.e. size and experimental settings, we have a total of \(10^5\) samples, with each sample containing a total of \(N = L \times L\) measurements. 
    }
    \centering
    \begin{tabular}{c|cccccc}
        \toprule
        Lattice size, $L$ & $5$ & $6$ & $11$ & $12$ & $15$ & $16$ \\
        \midrule
        Tokens ($10^8$) & $1.5$ & $2.16$ & $7.26$ & $8.64$ & $13.5$ & $15.4$ \\
        \bottomrule
    \end{tabular}
    \label{tab:measurements_lattice_sizes}
\end{table}

\begin{table}
    \caption{
    Details on the training of each model, including epochs trained, the amount of tokens in the training dataset, in units of \(10^8\), and the total number of tokens processed during training, in units of \(10^{10}\).
    The models were each trained for a constant amount of time, 85 hours, on an Nvidia A100.
    }
    \centering
    \begin{tabular}{c|c|c|c}
        \toprule
        Model & $M_1$ & $M_2$ & $M_3$ \\
        \midrule
        Epochs & 116 & 40 & 17 \\
        Tokens in dataset (\(10^8\)) & 3.66 & 19.6 & 48.4 \\
        Tokens processed (\(10^{10}\)) & 4.25 & 7.82 & 8.23 \\
        \bottomrule
    \end{tabular}
    \label{tab:training_details}
\end{table}
We conducted training on a single A100 GPU for 85 hours.
The training datasets for all three models are of the order of $10^8$, while during training, all models processed tokens on the order of $10^{10}$.
We summarize the relevant training information in \tabref{tab:training_details}.

\section{Inference with trained models}

In this section, we perform inference using the trained model and study the disordered-to-checkerboard phase transition of the Rydberg atom array, governed by the dimensionless detuning $\delta / \Omega$.
First, samples are drawn from the model in accordance to the method described at the end of Section~\ref{sec:nn_wavefunction}.
The samples are then used to obtain predictions of physical observables.
We compare our model's predictions with estimates obtained from QMC simulations~\cite{Merali2023StochasticSeriesExpansiona}.
For Rydberg atom arrays, QMC is able to obtain an estimate of the observables that can be considered exact, within statistical errors.
In showing that our model's predictions of these observables are in good agreement with the QMC values, we illustrate the efficacy of data-driven training as a technique for predicting Rydberg atom array measurement outcomes for ground states.
Further, we explore our trained models' ability to generalize away from the ground state and also to system sizes not represented in the training data. 

In order to construct our estimators below, we draw samples from our model decoder.
In our implementation, the complexity of sampling with respect to the system size $N = L \times L$ is of order $\mathcal{O}(N^3)$.
However, this complexity can theoretically be reduced to $\mathcal{O}(N^2)$ by caching the appropriate intermediate values in the attention calculation.

\subsection{Physical observables}
\label{sec:physical_observables}

Physical quantities that are diagonal in the occupation basis can be directly computed from the samples drawn from the model.
One such quantity is the staggered magnetization for the square-lattice Rydberg atom array.
The staggered magnetization is significant as it is the order parameter for the disorder-to-checkerboard quantum phase transition~\cite{Samajdar2020}, and it can be calculated with:
\begin{align}
    \langle\hat{\sigma}^{stag}\rangle  \approx \frac{1}{N_s} \sum_{\boldsymbol{\sigma} \sim p_\theta(\boldsymbol{\sigma})}
    \frac{1}{N}\left|   \sum_{i=1}^{N} (-1)^i  \left(n_i(\boldsymbol{\sigma}) - 1/2\right) \right| ,
\end{align}
where $i$ runs over all $N = L \times L$ atoms
using the
snake path convention of Fig.~\ref{fig:model_architecture}.
Here, 
$n_i(\boldsymbol{\sigma}) = \langle \boldsymbol{\sigma}| r_i \rangle\langle r_i|\boldsymbol{\sigma} \rangle$ is the occupation number operator acting on atom $i$ in a given configuration $\boldsymbol{\sigma}$. 

In contrast, we consider an off-diagonal observable, where we must make use of the ground state wave function amplitudes of the inferred samples $\Psi(\boldsymbol{\sigma}) = \sqrt{p_{\theta}(\boldsymbol{\sigma})}$. 
As an example, we examine the spatially averaged expectation value of $\hat{\sigma}^x$, which is defined as
\begin{align}
    \langle \hat{\sigma}^x \rangle & \approx \frac{1}{N_s} \sum_{\boldsymbol{\sigma} \sim p_\theta(\boldsymbol{\sigma})} \frac{1}{N}
    \sum_{\boldsymbol{\sigma}' \in \mathrm{SSF}(\boldsymbol{\sigma})} \frac{\Psi_\theta(\boldsymbol{\sigma}')}{\Psi_\theta(\boldsymbol{\sigma})}, \label{offdiagestimate} 
\end{align}
where $\mathrm{SSF}(\boldsymbol{\sigma})$ is the set of configurations that are connected to $\boldsymbol{\sigma}$ by a single spin flip (SSF).

Finally, we consider an estimate of the ground state energy $\langle E \rangle$, which is defined as
\begin{align}
    \langle E \rangle  
    &\approx \frac{1}{N_s} \sum_{\boldsymbol{\sigma} \sim p_{\theta}(\boldsymbol{\sigma})} \frac{\langle \boldsymbol{\sigma}|\Hat{H}|\Psi_{\theta}\rangle}{\langle \boldsymbol{\sigma}|\Psi_{\theta}\rangle}.
    \label{energy_estimate}
\end{align}
This observable can be efficiently evaluated for local non-diagonal operators~\cite{Carleo2017, Torlai2018Neuralnetworkquantumstate, HibatAllah2020Recurrentneuralnetwork, Melko2019RestrictedBoltzmannmachines}. More detailed derivations of  \eqref{offdiagestimate} and \eqref{energy_estimate} can be found in Appendix~\ref{app:offdiag}.

It is important to emphasize that these estimates of off-diagonal observables are appropriate only when the output of the decoder can be interpreted as the normalized wavefunction amplitude, $\Psi_\theta(\boldsymbol{\sigma}) = \sqrt{p_\theta(\boldsymbol{\sigma})}$. In particular, this correspondence is valid when the system is a pure state and the wavefunction can be assumed to be real and positive in the occupation basis, as is the case for the Rydberg Hamiltonian in its $T=0$ ground state. Importantly, the use of the decoder output as a wave function amplitude is no longer valid for finite temperature states. While the above estimates can still be computed, $\Psi_\theta(\boldsymbol{\sigma})$ is not able to capture the physics of the thermal state, as is demonstrated in our results which follow.

\begin{figure}
    \centering
    \includegraphics[width=0.48\textwidth]
    {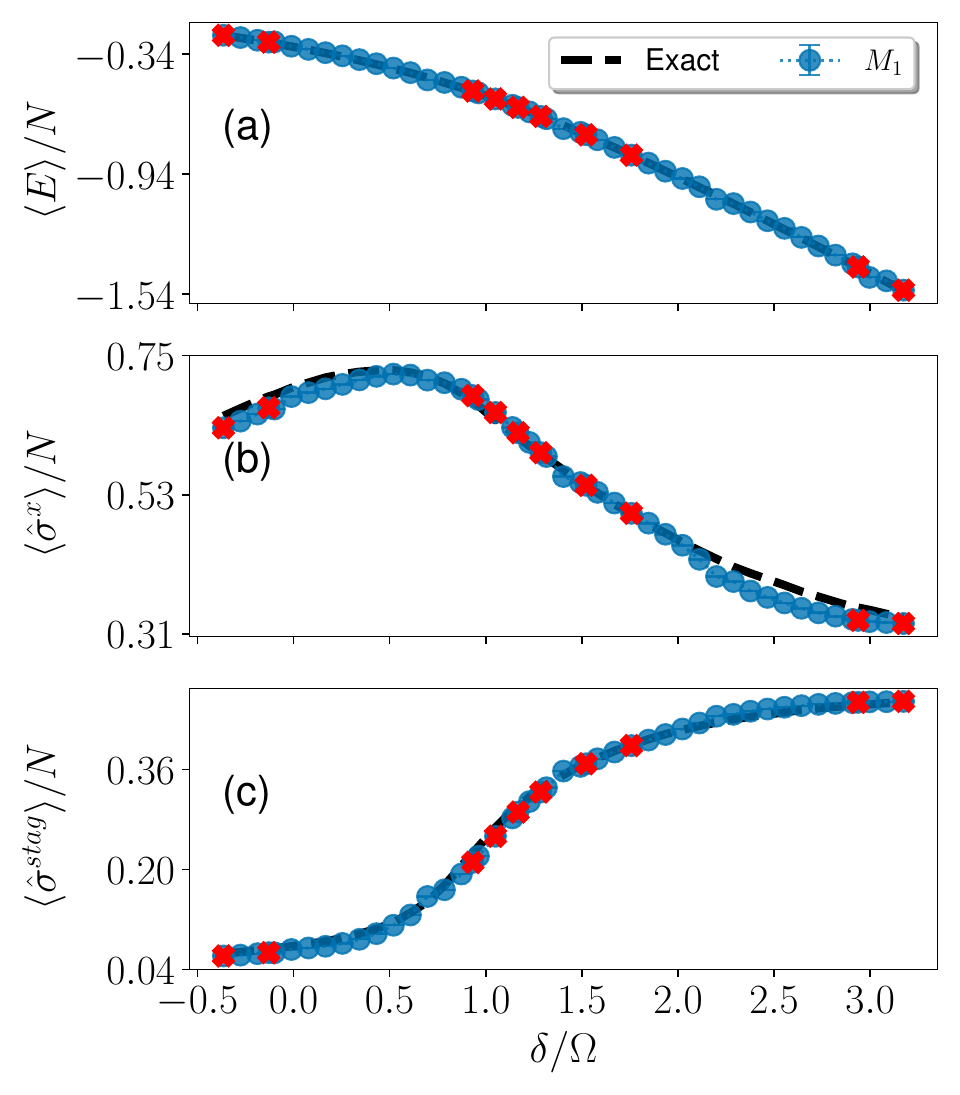}
    \caption{
    Model predictions of multiple observables across the disordered-to-checkerboard phase transition, which is governed by the dimensionless detuning, $\delta / \Omega$.
    The observables include, (a) energy, (b) site-averaged $x$-magnetization and (c) the staggered magnetization.
    Here, blue points correspond to observables estimated from projective measurements sampled from the trained model (i.e., out-of-distribution), red points indicate parameters contained within the training dataset (i.e., in-distribution), and dotted black lines correspond to the exact observables. 
    }
    \label{fig:estimators_vs_delta}
\end{figure}

\subsection{Results}

We now demonstrate the results of the trained models across parameter regimes and observables.
In \figref{fig:estimators_vs_delta} we illustrate the performance of model $M_1$ on the three observables introduced in Section \ref{sec:physical_observables} as a function of $\delta/\Omega$, for $R_b/ a = 1.15, \beta \Omega = 16.0$.
The estimators are calculated over $N_s=10^5$ samples.
The predictions are benchmarked against the QMC observables calculated from $10^5$ decorrelated samples, which we take as the ground truth.
At $\beta \Omega = 16$, the temperature is sufficiently low for the Rydberg atom array to be in its groundstate to a good approximation.
The model shows good agreement with the QMC estimates for the energy and staggered magnetization.
As for the $x$-magnetization, we observe a slight deviation from the QMC estimates in the tails.
We conclude that the trained model is able to generalize well to dimensionless detunings, $\delta / \Omega$, that lie outside of the training data set, when the system is restricted to the groundstate.

\begin{figure}
    \centering
     \includegraphics[width=0.48\textwidth]{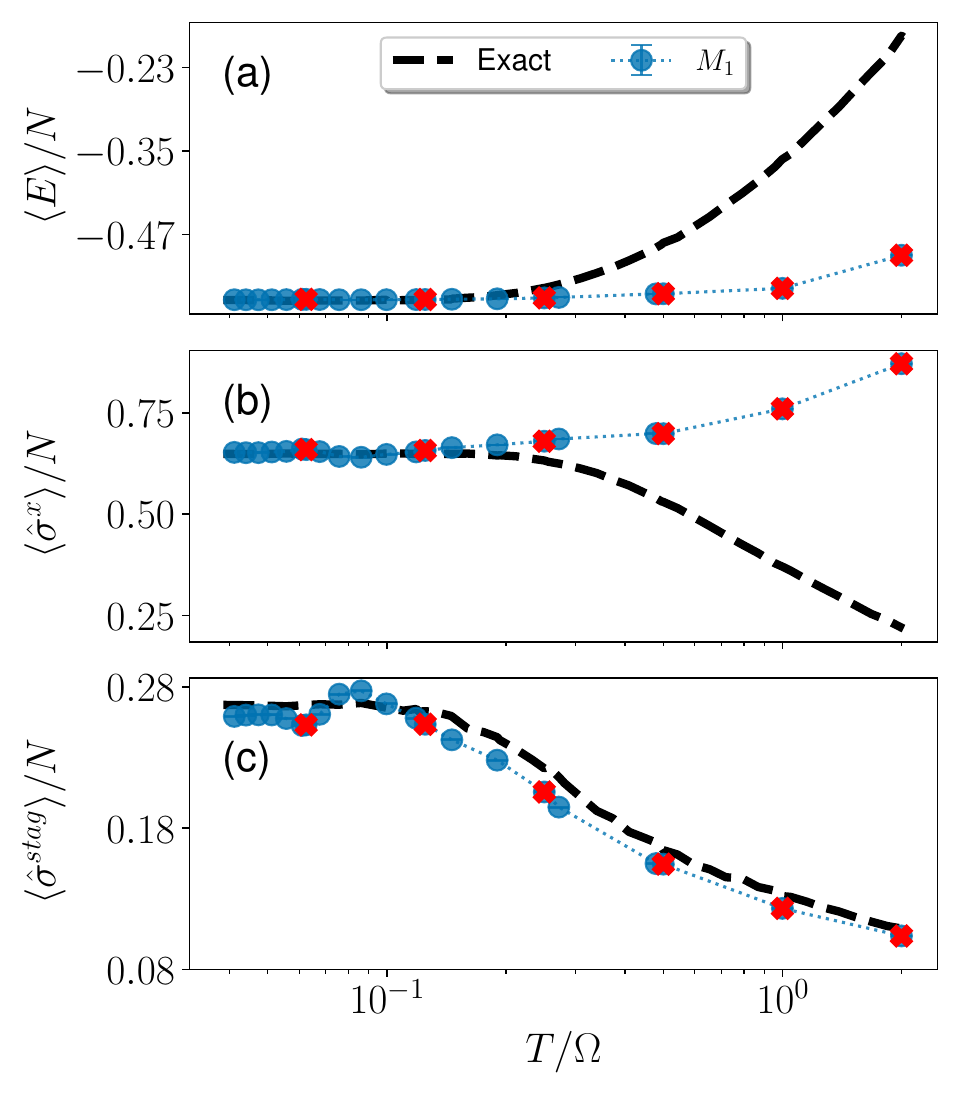}
    \caption{
    Model predictions of multiple observables as the dimensionless temperature $T / \Omega$ is increased.
    The observables include, (a) energy, (b) site-averaged $x$-magnetization and (c) the staggered magnetization.
    Again, blue points correspond to observables estimated from projective measurements sampled from the trained model (i.e., out-of-distribution), red points indicate parameters contained within the training dataset (i.e., in-distribution), and dotted black lines correspond to the exact observables. 
    }
    \label{fig:sigmax_vs_beta}
\end{figure}

Next, we investigate how well the model generalizes away from the groundstate.
In Fig.~\ref{fig:sigmax_vs_beta} we illustrate predictions made by model $M_1$ with the same strategy, except temperature $T = 1/\beta$ is varied instead of the detuning.
We see a strong disagreement between the exact observables and the model-obtained estimates at high temperatures.
In our architecture, the decoder represents a probability distribution over configurations in the Rydberg occupation basis, which can be mapped one-to-one onto a positive wavefunction of a pure state.
As mentioned, the $T=0$ ground state of the Rydberg Hamiltonian \eqref{eq:rydberg_hamiltonian} is known to take such a form, and thus the model performs well for the ground state, e.g. low temperatures.
However, as the temperature is increased, the system becomes a well-mixed Gibbs state, which can no longer be accurately represented with a positive wavefunction.
As such, the output of the decoder is incapable of accurately representing this state.
The model is merely approximating a probability distribution that results in identical occupation-basis measurement likelihoods.
This results in the incorrect estimate of off-diagonal observables, 
which require the wave function interpretation of the decoder output $\Psi_\theta(\boldsymbol{\sigma}) = \sqrt{p_{\theta}(\boldsymbol{\sigma})}$, as in \eqref{offdiagestimate}.
To appropriately capture such high temperature states, one might employ more sophisticated methods such as a complex density matrix \cite{kothe2023liouville} or minimally entangled typical thermal states (METTS) \cite{METTS1,METTS2}.

This is particularly evident in Fig.~\ref{fig:sigmax_vs_beta} at the limit of large temperature.
Here, the system tends to the fully-mixed state where all states are equi-probable.
The positive wavefunction that shares the same computational basis probability as this state is the equi-superposition of all states.
This state corresponds to the maximum eigenstate of the x-magnetization, leading to the divergence in Fig.~\ref{fig:sigmax_vs_beta} between the exact QMC and the trained model estimators at high temperatures, where $\langle \hat{\sigma}^x \rangle$ becomes significantly over-estimated. This is further supported by the fact that the high-temperature estimates of the staggered magnetization are somewhat accurate.
Again, this behavior is due to the fact that this diagonal estimator relies only on the computational basis probabilities with which certain configurations are sampled. 

\begin{figure}
  \centering
  \includegraphics[width=0.48\textwidth]{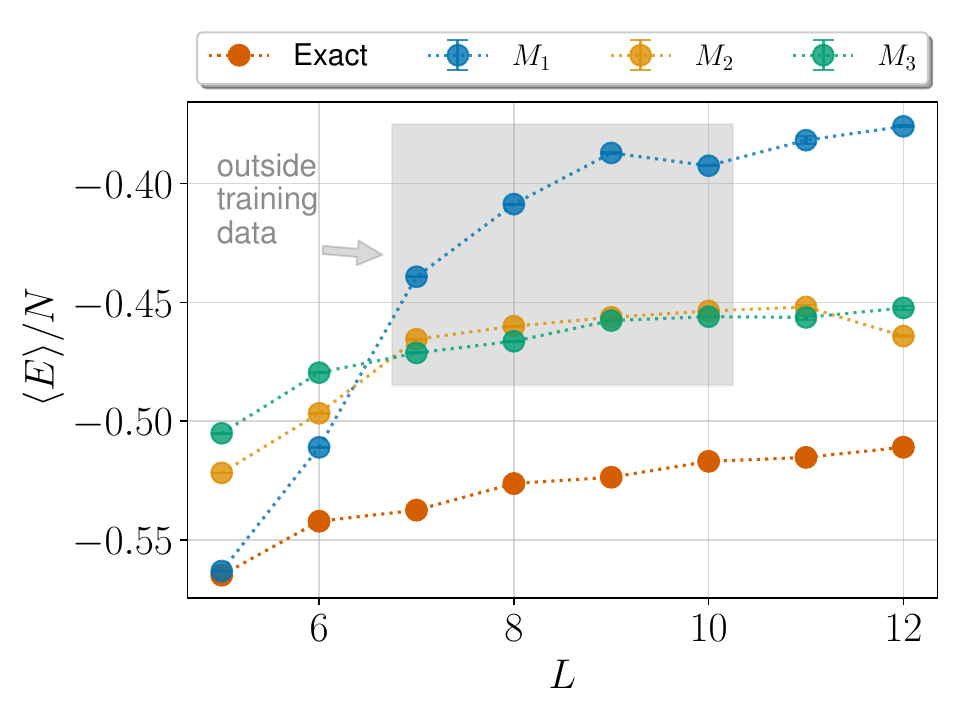}
    \caption{Extrapolation over lattice size $L$.
    The ground-state energy, as estimated by quantum Monte Carlo simulations and the models trained on three different datasets, demonstrates how training data affect the ability to extrapolate to system sizes not included in the training.
    }
  \label{fig:energy_vs_L}
\end{figure}

Finally, we assess the ability of our transformer to generalize to larger system sizes beyond those included in the training set.
For this we use all three trained models, $M_1$, $M_2$ and $M_3$.
Figure~\ref{fig:energy_vs_L} presents the energy evaluations from each of our three trained models, averaged over $10^4$ samples.
These models were trained on consecutively larger datasets as previously described in \tabref{tab:training_details}.
We first observe that the model $M_1$ trained on the smallest dataset, containing only data from lattice sizes $L = 5, 6$, captures the $L=5$ energy accurately but struggles to generalize to larger system sizes.
Interestingly, the models trained on more extensive datasets do not accurately capture the properties of their respective training sets, with the energy estimates deviating significantly from expected values.

These results illustrate how physical estimators are affected by bias in the training datasets.
As shown in~\tabref{tab:measurements_lattice_sizes}, we always accumulate $10^5$ samples for each grid size.
Therefore, the number of training tokens or measurements varies for each lattice size; more precisely, the dataset is biased towards larger lattice sizes.
This explains why models $M_2$ and $M_3$ perform worse than $M_1$ for the lattice sizes $L = 5, 6$, since the fraction of measurements for smaller sizes is less represented in their training dataset.
However, they perform better for larger grid sizes, where a larger relative number of training tokens are available.

\section{Conclusion}

In this paper we have explored the training of attention-based transformer models for the problem of mapping a quantum Hamiltonian to the qubit measurement outcomes of the corresponding quantum state.
We construct an encoder-decoder architecture suitable for Rydberg atom arrays, where the encoder input is a representation of the interacting Hamiltonian, and the decoder output represents measurements of the Rydberg occupation.

Since transformers have been developed as autoregressive models for natural language processing, they naturally represent probability distributions, such as Born-rule-distributed measurement outcomes.
In the case where a quantum state can be assumed to be pure, real and positive, this probability distribution has a one-to-one mapping to the wavefunction, giving the output of the decoder a powerful interpretation.
In the groundstate of Rydberg atom arrays governed by Hamiltonian \eqref{eq:rydberg_hamiltonian}, this assumption is valid, and we use it to demonstrate that a trained transformer model can accurately reproduce physical estimators in this case.

We have further explored the behavior of the transformer model when the quantum state is no longer real and positive, i.e. in a thermally mixed state at higher temperature.
We clearly see the breakdown in our ability to interpret the output of the decoder as the corresponding quantum state, as the physical estimators significantly deviate from their exact values for large $T$.
However, we propose that our decoder architecture could be modified to represent a density matrix \cite{kothe2023liouville} or METTS \cite{METTS1, METTS2}, which would be able to represent noisy, thermal and dynamical states, extending the capabilities of the model.

Finally, we attempt to demonstrate the scaling potential of the encoder-decoder transformer.
Namely, whether the transformer model, when trained on a dataset composed of smaller many-body system, is able to make predictions for larger system sizes.
Our results shows signatures of generalization, however, it also suggest that we have not attained the necessary scale in our transformer to exploit its full potential.
Hopefully, by training models with more data on a diverse set of lattice sizes, systematic improvements in model performance will be observed.
Indeed, such systematic scaling is the hallmark of large transformers used in industry, giving reason for optimism~\cite{kaplan2020scaling}.

More generally, just like in modern LLMs, our transformer architecture should allow for vastly improved performance with increases in the number of trainable parameters $\theta$, larger data sets, and more computational resources. 
In addition, since these models are being designed to represent physical qubit systems, a number of architectural improvements should be explored.
For example, there may be more efficient ways to use tokens than to represent a dictionary of only two individual qubit measurements.  
In particular, transformers designed for natural language processing typically manage dictionaries comprising tens to hundreds of thousands of tokens~\cite{Brown2020Languagemodelsare}.
An alternative approach involves using patches of qubits as tokens, which can reduce some computational demands~\cite{Sprague2024}.
However, this method introduces constraints on the system geometries, potentially limiting the applicability to arbitrary quantum array configurations.
This balance between computational efficiency and system flexibility remains a critical area for future research.

Notably, all three models discussed in this study were trained using only a single NVIDIA A100 GPU over a duration of 85 hours.
Looking forward, we can use this as a benchmark as the size and capability of such transformer models is increased towards the goal of predicting the behavior of larger quantum systems.
Inspired by the scaling observed in large language models within industry today, we imagine that significantly larger transformers trained on vastly more data may be able to make highly useful predictions, e.g.~about the behavior of a quantum computer in parameter regimes for which no data is currently available.

Finally, while our models in this study learned from synthetic data, our transformer architecture could instead be trained on qubit measurement data obtained from real experimental devices.
Such would be the beginning of a new type of {\it foundation model}, trained on real quantum computer output to generate predictions on a wide variety of unseen input settings.
Like today's LLMs, the capabilities of such large quantum models could increase dramatically with increased number of parameters, data set size, and compute, potentially ushering in a new era of scaling in AI and quantum computer co-design.

%STARTING POINT FOR A FOUNDATION MODEL

The RydbergGPT code and model weights for the models discussed in this study are available at~\cite{rydberggpt2024}.

\acknowledgments

We acknowledge support from the Natural Sciences and Engineering Research Council of Canada (NSERC) and the Perimeter Institute for Theoretical Physics. This research was supported in part by grant NSF PHY-2309135 to the Kavli Institute for Theoretical Physics (KITP). Computational support was provided by the facilities of the Shared Hierarchical Academic Research Computing Network (SHARCNET:www.sharcnet.ca) and Compute/Calcul Canada. 
Research at Perimeter Institute is supported in part by the Government of Canada through the Department of Innovation, Science and Economic Development Canada and by the Province of Ontario through the Ministry of Economic Development, Job Creation and Trade. D.F. acknowledges the Knut and Alice Wallenberg (KAW) Foundation for funding through the Wallenberg Centre for Quantum Technology (WACQT). The computations were enabled by resources provided by Chalmers e-Commons at Chalmers University of Technology. 

\bibliography{references}

\appendix

\section{Estimating expectation values of off-diagonal operators 
}\label{app:offdiag}

In this work, we estimate the ground state expectation values of two off-diagonal operators, $\langle\hat{\sigma}^x\rangle$ and $\langle E\rangle$, using samples inferred from our trained GPT model.
Here, we provide more detail for how to arrive at equations \eqref{offdiagestimate} and \eqref{energy_estimate}. 

Considering a general, off-diagonal operator $\hat{\mathcal{O}}$, we can write down its expectation value with respect to the probability distribution encoded in the trained GPT model,
\begin{equation*}
    \langle\hat{\mathcal{O}}\rangle
    = \frac{\langle \Psi_{\theta}\vert \hat{\mathcal{O}}\vert \Psi_{\theta}\rangle}{\langle \Psi_{\theta}\vert \Psi_{\theta}\rangle},
\end{equation*}
where $\Psi_{\theta}(\boldsymbol{\sigma}) = \sqrt{p_\theta(\boldsymbol{\sigma})}$ is the wave function interpretation of the joint probability distribution. 
Inserting the identity in the numerator and the denominator, and multiplying by 1, the above expectation value can be equivalently written as
\begin{equation*}
   \langle\hat{\mathcal{O}}\rangle = \frac{\sum_{\boldsymbol\sigma} \vert\langle \Psi_{\theta}\vert\boldsymbol\sigma\rangle\vert^2 \times \frac{\langle \boldsymbol\sigma\vert \hat{\mathcal{O}}\vert \Psi_{\theta}\rangle}{\langle \boldsymbol\sigma\vert \Psi_{\theta}\rangle}}{\sum_{\boldsymbol\sigma\prime} \vert\langle \Psi_{\theta}\vert\boldsymbol\sigma\prime \rangle \vert^2 }.
\end{equation*}
One can recognize 
$\frac{\vert\langle \Psi_{\theta} \vert\boldsymbol\sigma\rangle\vert^2}{\sum_{\boldsymbol\sigma\prime} \vert\langle \Psi_{\theta}\vert\boldsymbol\sigma\prime \rangle \vert^2}$
as some normalized probability $P(\boldsymbol\sigma)$ which can be approximated using importance sampling.
Here we also define the local estimator of our operator $\hat{\mathcal{O}}$,
\begin{equation*}
    \hat{\mathcal{O}}_{\text{loc}}(\boldsymbol\sigma) = \frac{\langle \boldsymbol\sigma\vert \hat{\mathcal{O}}\vert \Psi_{\theta}\rangle}{\langle \boldsymbol\sigma\vert \Psi_{\theta}\rangle},
\end{equation*}
such that the expectation value of $\hat{\mathcal{O}}$ can be estimated as
\begin{equation*}
    \langle \hat{\mathcal{O}}\rangle \approx \frac{1}{N_s}\sum_{\boldsymbol\sigma\sim p_{\theta}(\boldsymbol\sigma)} 
    \hat{\mathcal{O}}_{\text{loc}}(\boldsymbol\sigma).
\end{equation*}
If we take $\hat{\mathcal{O}} = \hat{H}$, then we have exactly recovered equation \eqref{energy_estimate}.

In order to arrive at equation \eqref{offdiagestimate}, we will consider what happens when the off-diagonal operator $\hat{\mathcal{O}}$ acts on the bra $\langle\boldsymbol\sigma\vert$ in the definition of $\hat{\mathcal{O}}_{\text{loc}}(\boldsymbol{\sigma})$.

If our off-diagonal operator $\hat{\mathcal{O}}$ is purely off-diagonal, it will map a given configuration configuration $\boldsymbol\sigma$ to a linear combination of some set of other configurations $\{\boldsymbol\sigma\prime\}$, that is
$\hat{\mathcal{O}}\vert\boldsymbol\sigma\rangle = \sum_{\{\boldsymbol{\sigma\prime}\}}
\text{C}_{\mathcal{O}}(\boldsymbol{\sigma\prime}) \vert \boldsymbol\sigma\prime\rangle$, and we can rewrite our expression for $\hat{\mathcal{O}}_{\text{loc}}(\boldsymbol{\sigma})$ as,
\begin{equation*}
    \hat{\mathcal{O}}_{\text{loc}}(\boldsymbol{\sigma}) = \sum_{\boldsymbol{\sigma}\prime \in \mathrm{SSF}(\boldsymbol{\sigma})}\text{C}_{\mathcal{O}}(\boldsymbol{\sigma}\prime)\frac{ \Psi_{\theta}(\boldsymbol{\sigma}\prime)}{\Psi_{\theta}(\boldsymbol\sigma)},
\end{equation*}
where we have used the fact that $\Psi_{\theta}(\boldsymbol\sigma) = \langle \boldsymbol\sigma\vert\Psi_{\theta}\rangle$.

We can note that if $\hat{\mathcal{O}} = \hat{\sigma}^x$, then $\mathrm{SSF}(\boldsymbol{\sigma})$ is the set of configurations connected to $\boldsymbol{\sigma}$ by a single spin flip (SSF) and $\text{C}_{\mathcal{O}}(\boldsymbol{\sigma}\prime) = 1$. Then it is clear that the inner sum in equation \eqref{offdiagestimate} takes the form of the equation above.

\end{document}